%% file: main.tex
\documentclass[sigconf]{acmart}

\AtBeginDocument{%
  \providecommand\BibTeX{{%
    \normalfont B\kern-0.5em{\scshape i\kern-0.25em b}\kern-0.8em\TeX}}}

\copyrightyear{2022}
\acmYear{2022}
\setcopyright{acmcopyright}\acmConference[SenSys '22]{The 20th ACM
Conference on Embedded Networked Sensor Systems}{November 6--9,
2022}{Boston, MA, USA}
\acmBooktitle{The 20th ACM Conference on Embedded Networked Sensor
Systems (SenSys '22), November 6--9, 2022, Boston, MA, USA}
\acmPrice{15.00}
\acmDOI{10.1145/3560905.3568435}
\acmISBN{978-1-4503-9886-2/22/11}

\begin{document}

\title[GaitVibe+: Enhancing Structural Vibration-based Footstep Localization with Temporary Cameras for Gait Analysis]{GaitVibe+: Enhancing Structural Vibration-based Footstep Localization Using Temporary Cameras for In-home Gait Analysis}
\providecommand\name{\textit{GaitVibe+}}

\author{Yiwen Dong}
\email{ywdong@stanford.edu}
\orcid{0000-0002-7877-1783}
\authornotemark[1]
\affiliation{%
  \institution{Stanford University}
  \streetaddress{450 Serra Mall}
  \city{Stanford}
  \state{California}
  \country{USA}
  \postcode{94305}
}
\author{Jingxiao Liu}
\orcid{0000-0001-5559-1627}
\affiliation{%
  \institution{Stanford University}
    \city{Stanford}
  \state{California}
  \country{USA}}

 \author{Hae Young Noh}
 \orcid{0000-0002-7998-3657}
\affiliation{%
  \institution{Stanford University}
    \city{Stanford}
  \state{California}
  \country{USA}}

\renewcommand{\shortauthors}{Dong, et al.}


\begin{abstract}
In-home gait analysis is important for providing early diagnosis and adaptive treatments for individuals with gait disorders.
Existing systems include wearables and pressure mats, but they have limited scalability due to dense deployment and device carrying/charging requirements. Recently, vision-based systems have been developed to enable scalable, accurate in-home gait analysis, but it faces privacy concerns due to the exposure of people's appearances and daily activities. To overcome these limitations, our prior work developed footstep-induced structural vibration sensing for in-home gait monitoring, which is device-free, wide-ranged, and perceived as more privacy-friendly. Although it has succeeded in temporal parameter estimation, it shows limited performance for spatial gait parameter estimation due to the low accuracy in footstep localization. In particular, the localization error mainly comes from the estimation error of the wave arrival time at the vibration sensors and its error propagation to wave velocity estimations. To this end, we present \name{}, a vibration-based footstep localization method fused with temporarily installed cameras for in-home gait analysis. Our method has two stages: fusion and operating stages. In the fusion stage, both cameras and vibration sensors are installed to record only a few trials of the subject's footstep data, through which we characterize the uncertainty in wave arrival time and model the wave velocity profiles for the given structure. In the operating stage, we remove the camera to preserve privacy at home. The footstep localization is conducted by estimating the time difference of arrival (TDoA) over multiple vibration sensors, whose accuracy is improved through the reduced uncertainty and velocity modeling during the fusion stage. We evaluate \name{} through a real-world experiment with 50 walking trials. With only 3 trials of multi-modal fusion, our approach has an average localization error of 0.22 meters, which reduces the spatial gait parameter error by 4.1x (from 111.4\% to 27.1\%) compared to the existing work.

\end{abstract}



\maketitle

\input{Sections/01Introduction}

\input{Sections/02Characterization.tex}
\input{Sections/03Method.tex}
\input{Sections/04Evaluation}
\input{Sections/05Conclusion}


\begin{acks}
This work was funded by the U.S. National Science Foundation (under grant number NSF-CMMI-2026699). The views and conclusions contained here are those of the authors and should not be interpreted as necessarily representing the official policies or endorsements, either express or implied, of any University, the National Science Foundation, or the United States Government or any of its agencies.
\end{acks}

\bibliographystyle{ACM-Reference-Format}
\bibliography{reference}

\end{document}

%% file: Sections/01Introduction.tex
\section{Introduction}\label{sec:intro}

In-home gait analysis for assessing the functional ability of a person's walking patterns is critical in providing early diagnosis and adaptive treatments for patients with neuroskeletal or neuromuscular disorders~\cite{lauziere2014understanding,pistacchi2017gait}. Many previous studies found that estimating spatio-temporal parameters such as step length, width, and duration for gait analysis is effective in predicting falls, functional loss, and mortality~\cite{doheny2010single,givon2009gait,bridenbaugh2011laboratory}. Quantitative measurements of gait health can help individuals understand their health status and disease risks, leading to improved life quality.

Existing gait analysis relies on direct observations and specialized sensing systems in well-calibrated gait clinics, but they are not suitable for continuous monitoring in non-clinical settings (e.g., in-home gait monitoring) due to the high cost of the equipment, installation constraints, and lack of professional staff. Other studies proposed wearables and pressure mats for in-home gait monitoring, but they have scalability limitations, such as having to carry a device or requiring dense sensor deployments. Recent studies developed camera-based gait monitoring, which is portable, cost-efficient, and accurate, but it has direct line-of-sight requirement and raises significant privacy concerns when installed at home for a long duration because the cameras also capture the owner's appearance and daily activities. To overcome these limitations, our prior works developed footstep-induced floor vibration sensing for gait health monitoring~\cite{Fagert2019,hahm2022home,dong2020md,rohal2022autoloc}, which is device-free, wide-ranged, and perceived as more privacy-friendly. Therefore, it is suitable for continuous in-home gait analysis. 



Although vibration-based in-home gait health monitoring has succeeded in temporal gait parameter estimation and gait symptom/balance characterization~\cite{dong2020md,jfarget2021}, it has limited performance in footstep localization, leading to inaccurate spatial parameter estimation. Previous studies developed localization methods mainly based on the time difference of arrival (TDoA) at multiple vibration sensors~\cite{mirshekari2018occupant,mirshekari2021obstruction,drira2021increasing,drira2019model,mirshekari2016non,pan2016multiple,mirshekari2016characterizing,shi2019device}. The average localization error of such methods is about 0.4 m, which is not accurate enough for gait analysis because the average human step length is $\sim$0.5 m. The main reason for the large error is that the existing methods overlook 1) \textit{the uncertainty in wave arrival time} (i.e., the time when vibration wave arrives at a sensor) and 2) \textit{the physical constraint of the wave traveling velocity over the structure}. Especially, they assume that their wave arrival time estimation is accurate when the wave velocity is optimized, which is often not true and would lead to a large error accumulation during footstep localization. In fact, the first challenge in vibration-based footstep localization is the uncertainty in wave arrival time. Unlike the impulse signal excited by a hammer strike that has an abrupt signal amplitude change, the initial arrival of the footstep-induced vibration wave has a relatively small amplitude due to the gradual increase in footstep forces (especially for patients), making it difficult to be detected out of the unpredictable noises in the sensing environment. Secondly, the wave propagation velocity is determined by the structural property of the floor, which varies over different locations of the structure. It is difficult to estimate the velocity without knowing the details of the structural properties.

To address these challenges, we present \name{}, a new vibration-based footstep localization method through multi-modal fusion with temporarily installed cameras for in-home gait analysis. Our system characterizes the uncertainty in wave arrival time and models the velocity profile of a given structure by collaboratively analyzing vision and vibration signals. \name{} consists of two stages: 1) fusion stage (vibration and vision), and 2) operating stage (vibration only). During the fusion stage, both cameras and vibration sensors are temporarily installed for a short period of time with the user's awareness and consent to record only 3-5 trials of the subject's footstep data. Such information is then translated into a wave velocity profile model and wave arrival time estimation. Then, cameras are removed during the operating stage to preserve privacy in individuals' homes for long-term monitoring. The footstep localization is conducted by vibration sensors based on TDoA estimation, whose accuracy is improved through the reduced uncertainty in wave arrival time and velocity modeling during the fusion stage. 

To evaluate our method, we conducted a real-world experiment with 50 walking trials on a wooden floored structure. With only 3 trials of vibration-vision fusion, we successfully reduce the average localization error to 0.22 meters, which reduces the error from 111.4\% to 27.1\% (by 4.1$\times$) for spatial parameter estimation.

The contributions of this paper are:
\begin{itemize}
\item We develop \name{}, a new vibration-based footstep localization system through multi-modal fusion with temporarily installed cameras for in-home gait analysis.
\item We characterize the uncertainty in wave arrival time and spatial variations in wave propagation velocity to reduce the uncertainty and model the velocity profile for a given structure through combined vibration and vision analysis, resulting in higher accuracy in vibration-based footstep localization.
\item We evaluate \name{} through a field experiment of 50 walking trials and successfully reduced the gait parameter error by 4.1 times (from 111.4\% to 27.1\%).
\end{itemize}

The remainder of the paper first presents the characterization of uncertainty in wave arrival time and variations in wave propagation velocity (Section~\ref{sec:characterization}). Then, we introduce our two-stage method for accurate spatial gait parameter estimation (Section~\ref{sec:method}). Finally, we discuss the real-world experiment and evaluation results (Section~\ref{sec:eval}), followed by future work (Section~\ref{sec:futurework}) and conclusions (Section~\ref{sec:conclusion}).

%% file: Sections/02Characterization.tex
\section{Characterizing the Wave Arrival Time Uncertainty and Velocity Variations for Multi-Modal Fusion} \label{sec:characterization}

We characterize the uncertainty in wave arrival time and variations in wave propagation velocity with the help of the cameras to understand the error sources in vibration-based footstep localization.

\begin{figure}[t!]
\begin{center}
    \includegraphics[width=\linewidth]{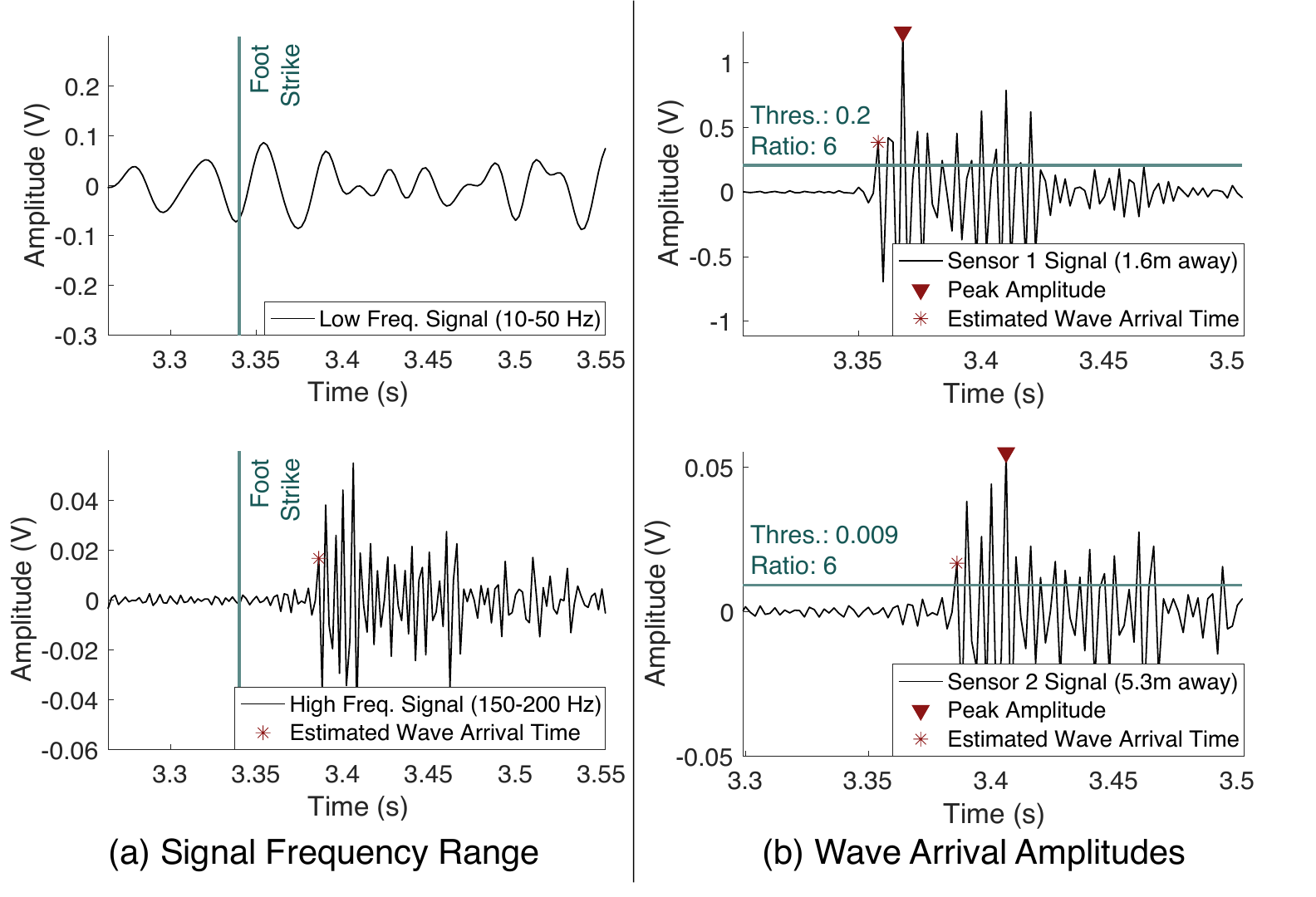}
  \caption{Characterization of wave arrival time uncertainty: (a) the high-frequency range has a clearer wave arrival peak than the low-frequency range; (b) different footstep-to-sensor distances require distinct thresholds for wave arrival amplitudes, but they share consistent ratios when compared to the peak amplitude.}%
\label{fig:wave_arrival}
\vspace{-15pt}
\end{center}
\end{figure}

\subsection{Uncertainty in Wave Arrival Time} \label{sec:arrival_char}

The arrival times of footstep-induced floor vibration waves are highly uncertain due to 1) the signal overlaps from the current and previous footsteps and 2) the relatively small signal amplitude at the beginning of each footstep. First, human gait cycles are typically continuous, meaning that the vibration wave generated by the previous footstep overlaps with the next one. It is difficult to accurately identify the arrival time of each footstep's vibration wave. In addition, unlike hammer trikes and other impulsive forces, footstep force gradually increases after the foot touches the ground, so the peak detection approach that works for impulsive events does not capture the initial arrival time of human footsteps accurately due to the presence of unpredictable noises from the environment.

To characterize the uncertainty in wave arrival time, we combine the knowledge of foot-strike time and location captured obtained from video and the gait characteristics reflected in the vibration signals. Such uncertainty is characterized by the following perspectives:

\textbf{a) Signal frequency range vs. foot-strike time:} We characterize the frequency distributions of a single-footstep signal vs. foot-strike time captured by the cameras to choose the frequency range for wave arrival detection. As shown in Figure~\ref{fig:wave_arrival}a, while the low-frequency range (10-50 Hz) has higher amplitudes in signals than the high-frequency range (150-200 Hz), the wave arrival peak is not obvious and has significant overlaps with the previous footstep. Therefore, we extract the high-frequency range to detect the arrival time.
    
\textbf{b) The signal amplitude at wave arrival vs. footstep-to-sensor distances:} To accurately estimate the vibration wave arrival time at the sensors despite its small amplitude, we compare the wave arrival amplitude (i.e., the signal amplitude upon wave arrival) vs. the wave propagation distances captured by the cameras. As shown in the y-axis scale of Figure~\ref{fig:wave_arrival}b, due to the wave attenuation effect, the sensor closer to the footstep observes a significantly higher wave arrival amplitude than the further away sensor. Therefore, detecting the wave arrival time using a constant threshold on amplitude leads to errors because of the varying sensing distances when the subject is walking. To address this problem, we leverage the insight that the wave attenuation rate is the same given the same propagation distance and frequency, which means that the signal amplitudes within the same frequency range at each sensor share the same attenuation rate. Therefore, the ratio between the wave arrival amplitude and the peak amplitude within that signal does not depend on the footstep-to-sensor distances. 
To this end, we compute this ratio for each sensor because this ratio is consistent over various sensing distances (see Figure~\ref{fig:wave_arrival}b).

\begin{figure}[t!]
\begin{center}
    \includegraphics[width=\linewidth]{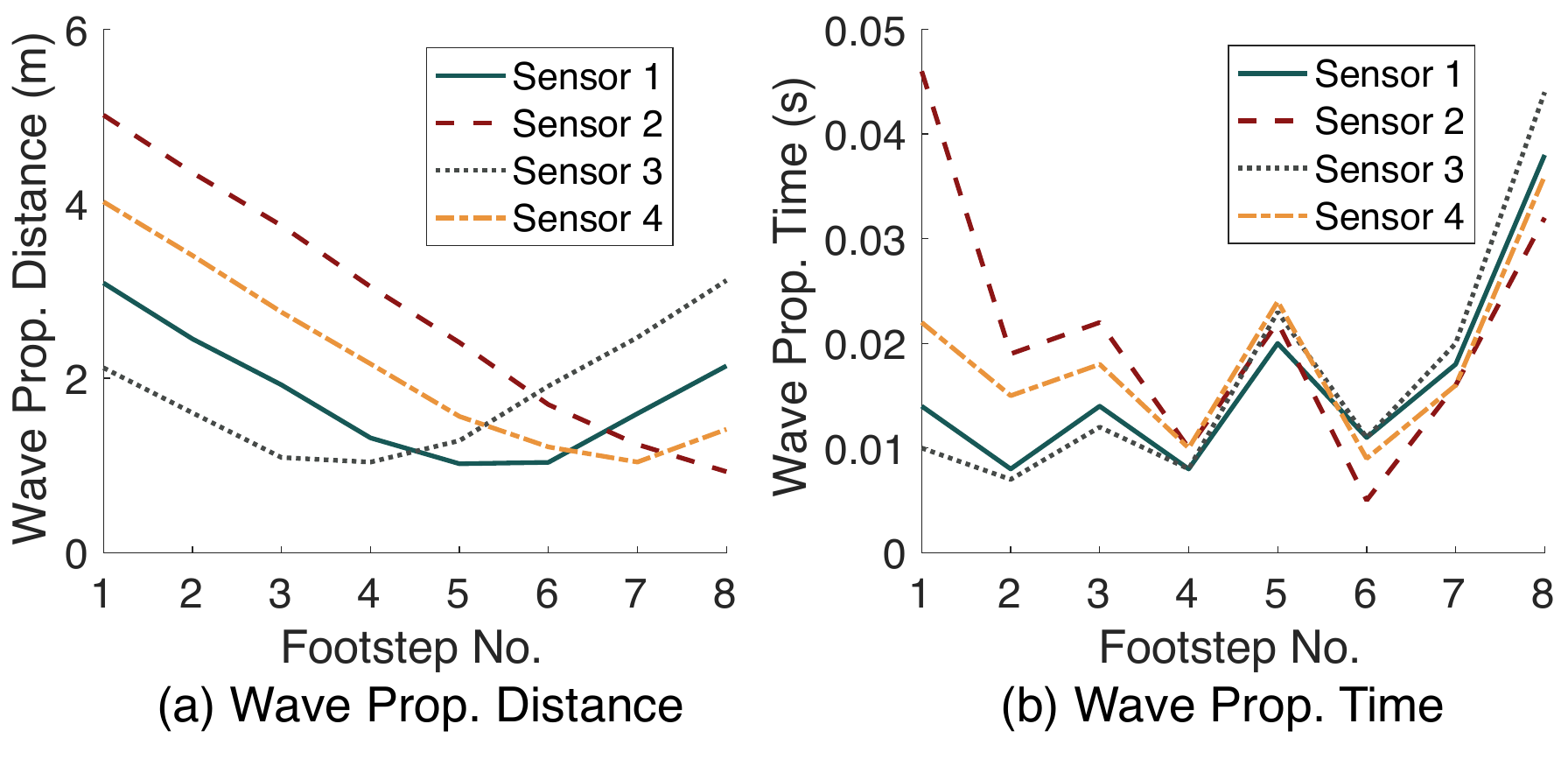}
  \caption{Variations in wave propagation velocity observed based on (a) wave propagation distance and (b) wave propagation time through vision- and vibration-based sensing. Distance and time have different trends, meaning that the velocities are different at different footstep locations.}%
\label{fig:wave_velocity}
\vspace{-10pt}
\end{center}
\end{figure}

\subsection{Variations in Wave Propagation Velocity}\label{sec:velocity_char}

The wave velocities of footstep-induced floor vibration vary across different locations of the floor because of the structural heterogeneity~\cite{mirshekari2018occupant}. The wave propagation velocity is affected by material properties, defects/cracks in the structure, and the type of connections between structural components. Therefore, when a person's footstep location changes, the underlying structural property changes, resulting in changes in the vibration wave velocity.

We characterize the wave propagation velocity by dividing the wave propagation distance by the propagation time. The wave propagation distance is captured by the cameras. As shown in Figure~\ref{fig:wave_velocity}a, the distances first decrease and then increase when a person passes by. Since the sensors are mounted at different locations, their wave propagation distance trends have a slight offset along the person's walking path. Besides, the wave propagation velocity is estimated by taking the time difference between the foot-strike time captured by the camera and the wave arrival time in the vibration signals. Figure~\ref{fig:wave_velocity}b shows the wave propagation time corresponding to the distance change described in Figure~\ref{fig:wave_velocity}a, which has an irregular shape compared to the distances. There are several findings about the velocity after dividing the distance by time for these four sensors in the figure: 

\textbf{a) Velocity variations across different locations of the structure:} We observe that the velocity of the vibration wave varies across the structure because the wave propagation distance and time have different trends (see Figure~\ref{fig:wave_velocity}). During preliminary testing of our structure, the calculated wave traveling velocity is between 30 m/s and 300 m/s. In order to estimate the wave velocity accurately for each footstep location, we model the velocity profile across the floor based on multiple walking trials during the fusion stage, which is discussed in greater detail in Section~\ref{sec:velocity_profile} with Figure~\ref{fig:velocity_profile}.

\textbf{b) Velocity variations across different directions of wave propagation:} We observe that the velocity of the vibration wave also varies across different directions. Since sensors 1-4 are mounted at the side of the walking path, their wave propagation directions for each footstep are different. For example, at the footstep location 5 shown in Figure~\ref{fig:wave_velocity}, the wave traveling velocity has a range of $\sim$70-130 m/s among these four sensors. Compared to the velocity range across different locations (30-300 m/s), the effect of different wave propagation directions is less significant. In addition, it is not realistic to estimate velocity in every direction for all footstep locations because there is an unlimited number of wave propagation directions at each location. Therefore, in this work, we assume that the velocity at each footstep location is the same for different directions. With this assumption, the order of wave arrivals across multiple sensors can be determined by the order of footstep-to-sensor distances.

%% file: Sections/03Method.tex
\section{Enhancing Vibration-based Footstep Localization
using Temporary Cameras }\label{sec:method}

\begin{figure}[t!]
\begin{center}
    \includegraphics[width=\linewidth]{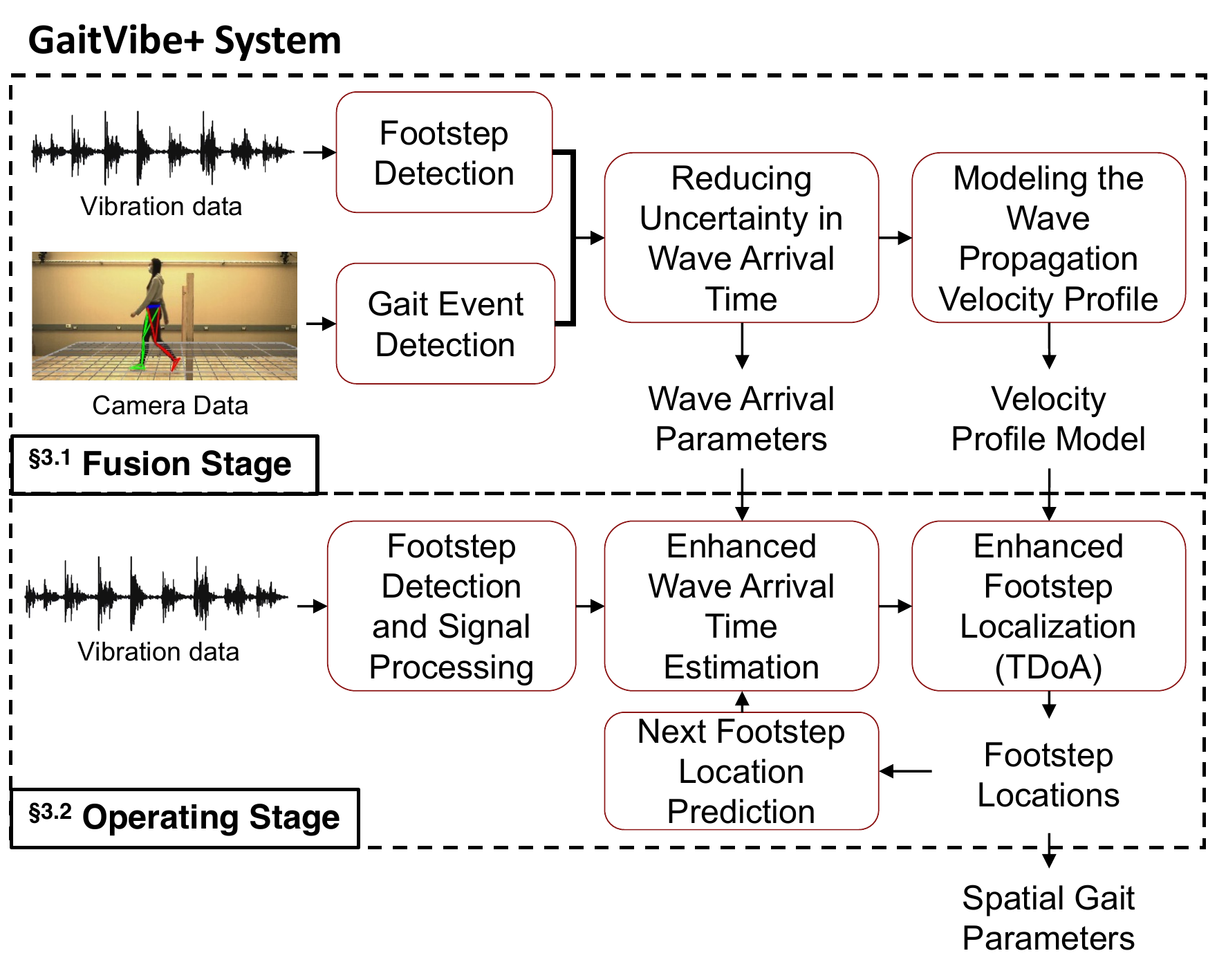}
  \caption{\name{} system consists of two stages: 1) fusion stage (vibration and vision), 2) operating stage (vibration). }%
\label{fig:system}
\vspace{-15pt}
\end{center}
\end{figure}

\name{} consists of two stages: 1) the fusion stage, where we combine vibration and vision to reduce the uncertainty in wave arrival time and model the velocity profile across various locations of the structure, and 2) The operating stage, where the cameras are removed, and the vibration-based system alone performs improved localization on the fly (see Figure~\ref{fig:system}). 

\subsection{Fusion Stage: Vibration and Vision Fusion for Localization Uncertainty Reduction}
During the fusion stage, temporary cameras are installed with vibration sensors to record a small number (e.g., 3$\sim$5) of walking trials from the subject when walking alone. By collaboratively analyzing the data from both modalities, we estimate parameters for reducing the uncertainty in wave arrival time and modeling the wave velocity profile of the given structure.

\subsubsection{Vibration and Vision Data Processing}
The first step in the fusion stage is to collect a few walking trials from both the cameras and the vibration sensors to prepare for data fusion.

To process the floor vibration data, we apply an adaptive Wiener filter to remove the white noise and then detect individual footsteps by computing the wavelet coefficients over the range of 0-300 Hz (a range that contains the most footstep-related information based on our prior studies~\cite{dong2023stranger,dong2020md}). Next, footsteps are detected by a peak-picking algorithm over the time domain and then segmented by a fixed window of 0.8 seconds (i.e., the maximum duration of a footstep in our data) around the detected peaks. 

The camera data is processed through a built-in software package for gait analysis~\cite{Limited2015}. First, the software reconstructs the moving trajectory of the heel in the 3D space. With these moving trajectories, we extract the time of the foot strike by detecting the minimum vertical position of the heel. The footstep location is obtained as the x- and y-axis coordinates in the 2D floor plane at the foot-strike time. 

\subsubsection{Reducing Uncertainty in Wave Arrival Time}
In this step, we extract parameters for wave arrival time estimation to reduce uncertainty and then estimate the wave arrival time using these parameters for velocity profile modeling. 

The parameters for wave arrival time estimation include 1) the frequency range of the signal, 2) the noise level of the sensors, and 3) the ratio between peak amplitude and wave arrival amplitudes. The frequency ranges of the signal are determined based on the characterization results in Section~\ref{sec:arrival_char}, where we choose the high-frequency range that shows a clear wave arrival time in the vibration signals. 
The noise level of each sensor is extracted when there is no footstep or disturbance from the environment. We estimate the mean and the standard deviation of the noise for each sensor in order to detect the footsteps and the initial wave arrival peaks during the operating stage. The ratio between the peak amplitude and the wave arrival amplitude is computed to reduce the uncertainty during the operating stage when the exact foot-strike time is unknown after the removal of the cameras. 

The wave arrival time is estimated through a combined analysis of vision and vibration data. There are two steps for estimation: 1) per-sensor wave arrival time estimation and 2) multi-sensor wave arrival sequence shortlisting. We first estimate the wave arrival time candidates for each sensor through peak-picking and anomaly detection algorithms on signals between the foot-strike time and the time when the peak amplitude occurs. This is because the footstep force gradually increases after the initial contact with the floor, so the range of wave arrival time is always between the foot-strike time and the peak amplitude time. Then, we combine the candidates from multiple sensors to compute the sequence of wave arrival across sensors. Since the sensor closer to the footstep typically receives the wave first (as discussed in Section~\ref{sec:velocity_char}), we further shortlist the arrival time combinations based on the order of footstep-to-sensor distances. Finally, we select the earliest wave arrival time combination observed in the vibration signal to allow an accurate estimation of the wave propagation velocity. 

\begin{figure}[t!]
\begin{center}
    \includegraphics[width=0.9\linewidth]{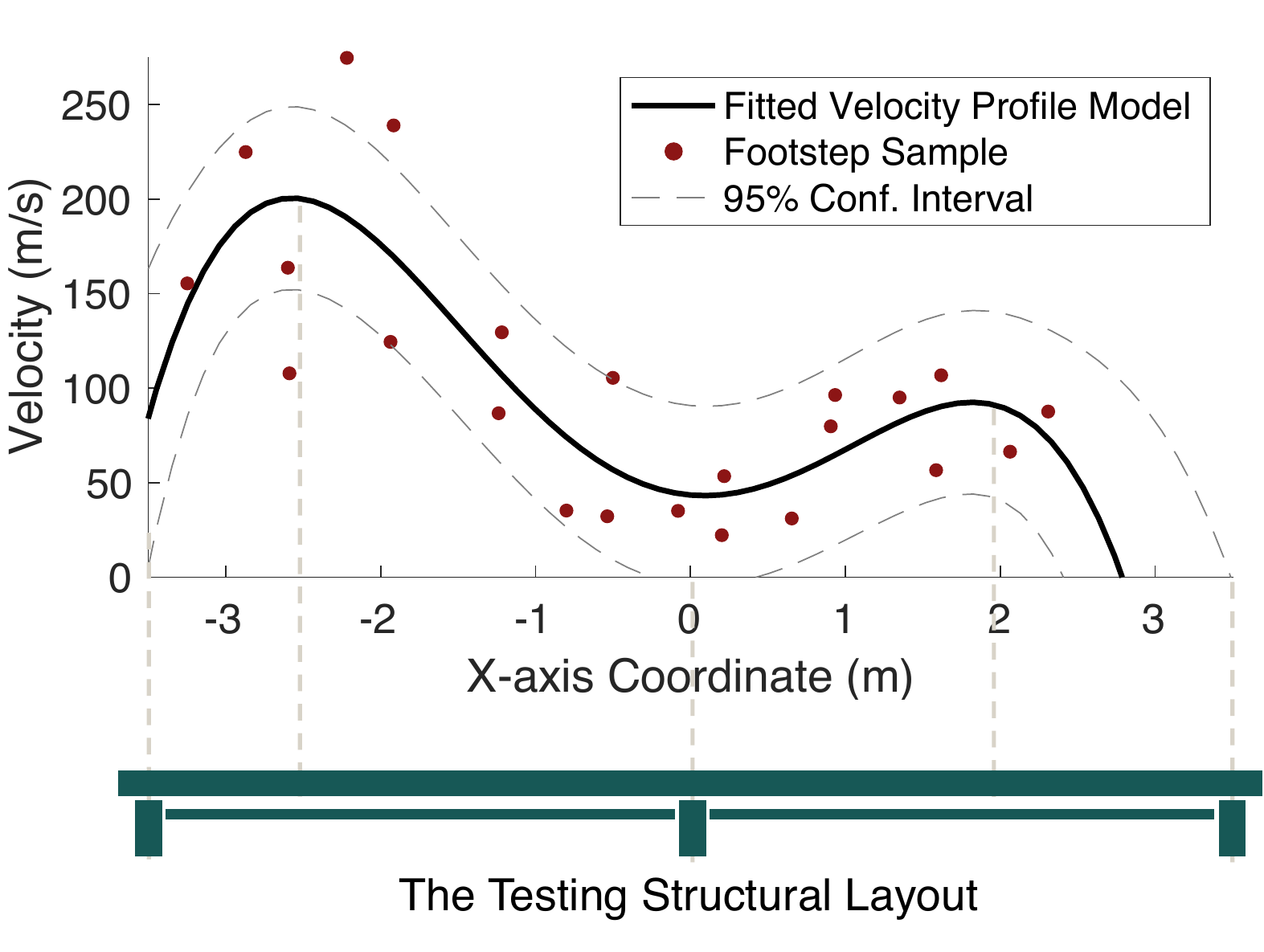}
  \caption{The velocity profile model for the testing structure. Overall, the velocity profile is symmetrically shaped, which corresponds well with the structural layout. The higher velocity on the left may be caused by deviations during the manual assembly of the structural components.}%
\label{fig:velocity_profile}
\vspace{-15pt}
\end{center}
\end{figure}

\subsubsection{Modeling the Wave Propagation Velocity Profile}\label{sec:velocity_profile}

We model the wave propagation velocity profile based on the wave propagation distance and time at various footstep locations observed by cameras and vibration sensors. As described in Section~\ref{sec:characterization}, the velocity at each footstep location is computed by distance over time, where the distance is calculated based on the footstep location captured by the cameras, and the time is obtained as the duration between the foot-strike time and the wave arrival time. Assuming the wave propagation velocity is consistent at a given footstep location, we model the velocity profile based on the calculated velocities over various locations of the structure.

To develop the velocity profile model, we conduct a non-linear regression on the footstep samples collected during the fusion stage to model the velocity profile over the area of coverage by the vibration sensors. The regression model we chose is a 4$^{th}$ order polynomial model because it describes the non-linear velocity trend on the structure and has a consistent training and validation accuracy during preliminary testing, meaning that it does not over-fit to the individual data samples that may reflect the local defects in the structure.

Figure~\ref{fig:velocity_profile} shows the velocity profile along the longitudinal center line of our testing structure. It is a wooden-framed structure that has two spans over the length and the same cross-section over the width. The estimated velocity profile corresponds well with its cross-section layout - the vibration wave appears to travel slower at locations with columns and faster at the mid-span of the structure.

\begin{figure}[t!]
\begin{center}
    \includegraphics[width=\linewidth]{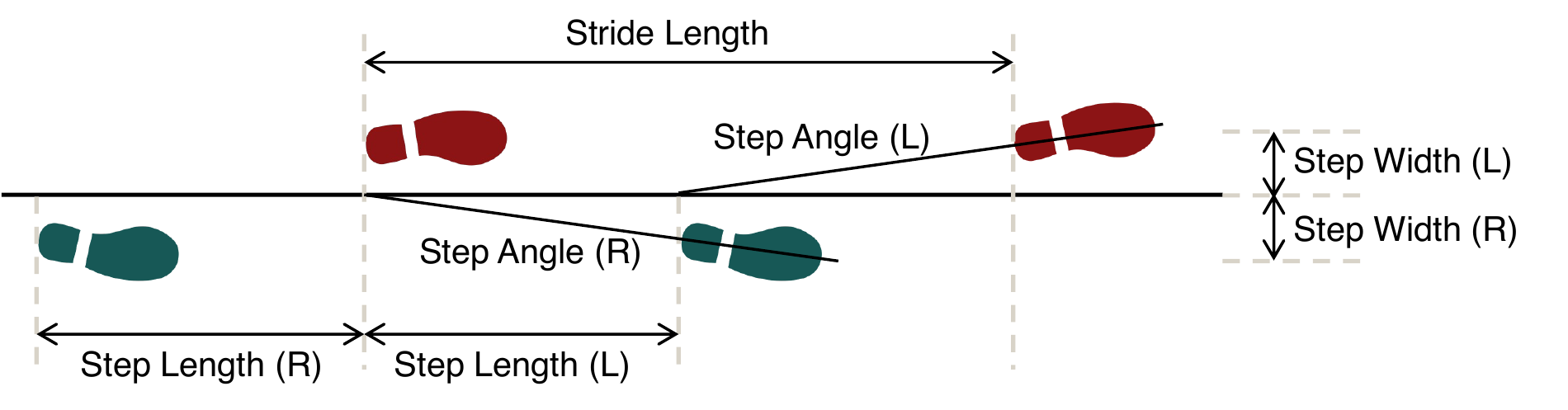}
  \caption{An illustration of how the spatial gait parameters are estimated based on the footstep locations.}%
\label{fig:spatial_param}
\vspace{-10pt}
\end{center}
\end{figure}

\subsection{Operating Stage: Structural Vibration-based Footstep Localization}
During the operating stage, cameras are removed to preserve privacy at home, and the vibration sensors alone localize footsteps with better accuracy based on the wave arrival parameters and velocity profile model. To achieve this, we first process the vibration signal to detect footsteps and predict the potential next footstep locations based on the previously observed footsteps. Then, we estimate the wave arrival time at each sensor based on the predicted next footstep location and the wave arrival parameters learned during the fusion stage. After that, we localize the footsteps through multi-lateration based on the velocity profile model and the time difference of arrival (TDoA) over multiple sensors. We choose the TDoA-based approach because it is less sensitive to the change of floor types and the shoe types~\cite{mirshekari2016characterizing,mirshekari2016non}, so it is more scalable to different people's homes. Finally, we compute the spatial gait parameters based on the estimated footstep location for in-home gait analysis.

\subsubsection{Footstep Detection and Signal Processing}
We process the vibration signals to detect footsteps and activate our \name{} system. Footsteps are detected by an anomaly detection algorithm~\cite{dong2023stranger} when the signal amplitude is above three standard deviations of the noise level. Then, an adaptive noise filter is applied to the signals. Similar to the fusion stage signal processing, individual footsteps are segmented through peak picking and window cropping around the detected peaks.  

\subsubsection{Next Footstep Location Prediction}
Since a human walking trial typically consists of a series of continuous footsteps, the location of each footstep depends on the previous footstep location. Therefore, we predict the potential next footstep locations based on the previously observed footstep locations. This location prediction is an area centered around the previous footstep location: the length and the width of the area is 1 meter, which is computed based on the maximum step length plus three standard deviations of its variance during preliminary testing among all participants. The location prediction of the first detected footstep is a 1-meter soft boundary around the sensing area.

\subsubsection{Enhanced Wave Arrival Time Estimation}
The uncertainty in estimating the arrival time of footstep-induced vibration waves is reduced based on three criteria: 1) the range of the arrival time, 2) the order of wave arrival over multiple sensors, and 3) the range of the wave arrival amplitude. First, the range of the arrival time is considered as the time between the signal exceeding the 90\% confidence interval of the noise level and the signal reaching the peak amplitude within a footstep. We detect peaks within that time range for each sensor and combine the results from multiple sensors to provide a list of wave arrival time combinations. Then, we shortlist the combinations according to the wave arrival orders based on the next footstep location prediction. To further reduce the uncertainty over these candidate combinations, we compute the range of wave arrival amplitude based on the ratio estimated during the fusion stage. The earliest combination that satisfies all criteria is estimated as the wave arrival time. 

\subsubsection{Enhanced Footstep Localization with the Velocity Profile Model}
The footstep is localized based on the velocity profile model and the time difference of arrival (TDoA) across multiple sensors. First, we estimate the range of wave propagation velocity using the velocity profile model and the footstep location proposal. Then, we compute TDoA over multiple sensors by subtracting the arrival time at the sensor with the largest signal amplitude. Finally, the footstep is localized through a grid search over the footstep location proposal to choose a velocity resulting in a TDoA that is the closest to the TDoA obtained from the vibration signals.

\subsubsection{Spatial Gait Parameter Estimation}
We estimate the gait parameters based on the predicted footstep location. These parameters include step length, stride length, step width, step angle, and walking speed. The meaning of these parameters is illustrated in Figure~\ref{fig:spatial_param}.

\begin{figure}[t!]
\begin{center}
    \includegraphics[width=0.8\linewidth]{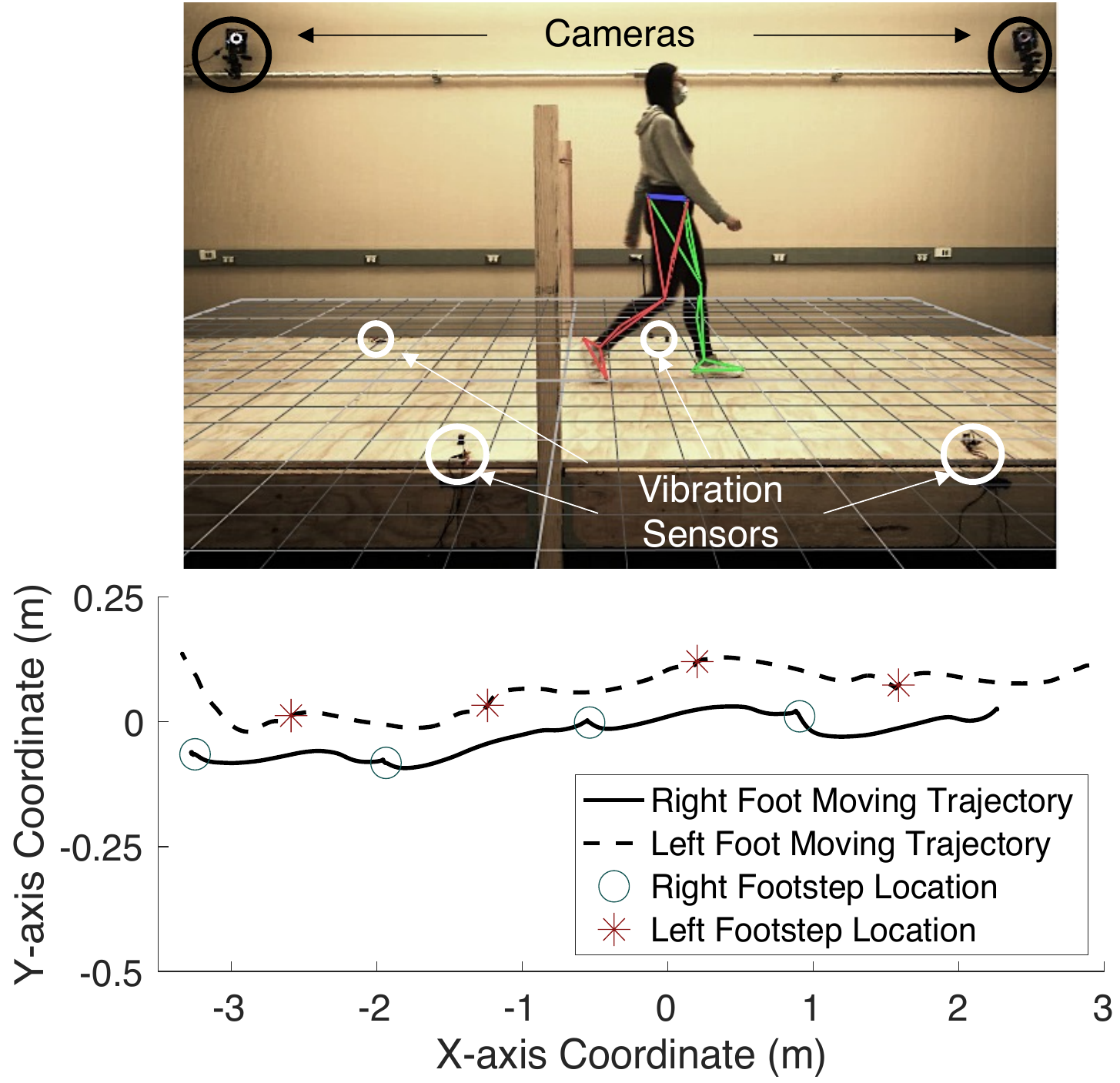}
  \caption{Experiment setup (upper figure) and a sample walking trial from a person extracted by analyzing the camera data (lower figure).}%
\label{fig:experiment}
\vspace{-10pt}
\end{center}
\end{figure}

%% file: Sections/04Evaluation.tex
\section{Real-world Evaluation} \label{sec:eval}
We evaluate \name{} through a real-world experiment with 50 walking trials on a wooden structure. In this section, we describe the experimental setup and discuss the performance of \name{} in terms of footstep localization and spatial gait parameter estimation.

\subsection{Experiment Setup}
As described in Figure~\ref{fig:experiment}, four SM-24 geophone (i.e., vibration sensors) with 500 Hz sampling frequency were attached to the floor~\cite{geophone}. For ground truth collection during the experiment, eight infrared cameras driven by the Vicon Motion Capture system for professional gait health assessments (100 Hz sampling frequency) were mounted on the rack below the ceiling to capture the moving trajectories of the subjects' on-body markers~\cite{Limited2015}. The number of cameras can be significantly reduced, and the type of cameras will be switched to normal video cameras for in-home deployments given the promising results of 3D pose estimation from video/RGB cameras~\cite{martinez2017simple,mehta2017vnect}, which will be incorporated in our future work. 

The geophones were mounted along the two edges of a 7-meter-long wooden-framed structure, spaced apart by 2 meters. Amplifiers were used to improve the signal-to-noise ratio (SNR), increasing the sensing range to up to 10 meters in diameter. A Vicon Lock Lab system was connected to the vibration sensors to acquire, synchronize and convert the analog signal to the digital signal~\cite{Limited2015}. The cameras from the Vicon Motion Capture system were mounted by tripods and clips attached to the hanging bars on the wall, aiming at the wooden structure~\cite{Limited2015}. 

The participants were asked to walk across the platform more than 20 times using their natural gait. For each walking trial, markers were attached to the subjects' lower limbers to capture the moving trajectories of their heels and toes to collect the ground truth for reference. Gait events of the first cycle for each person were labeled manually and then extrapolated through auto-correlation for the remaining cycles. All experiments were conducted in accordance with the approved IRBs (IRB-55372).

\subsection{Performance of \name{}}
The performance of \name{} is evaluated with 50 walking trials, including the initial 3 trials for the fusion stage and 47 trials for the operating stage. In this study, the number of trials for fusion is determined by the range limit ($\sim$100 m/s) of the confidence interval of the estimated wave velocity. Overall, our approach has achieved a 3$\times$ error reduction in footstep localization, which leads to a 4.1$\times$ error reduction in spatial parameter estimation (from 111.4\% to 27.1\%).

\begin{figure}[t!]
\begin{center}
    \includegraphics[width=\linewidth]{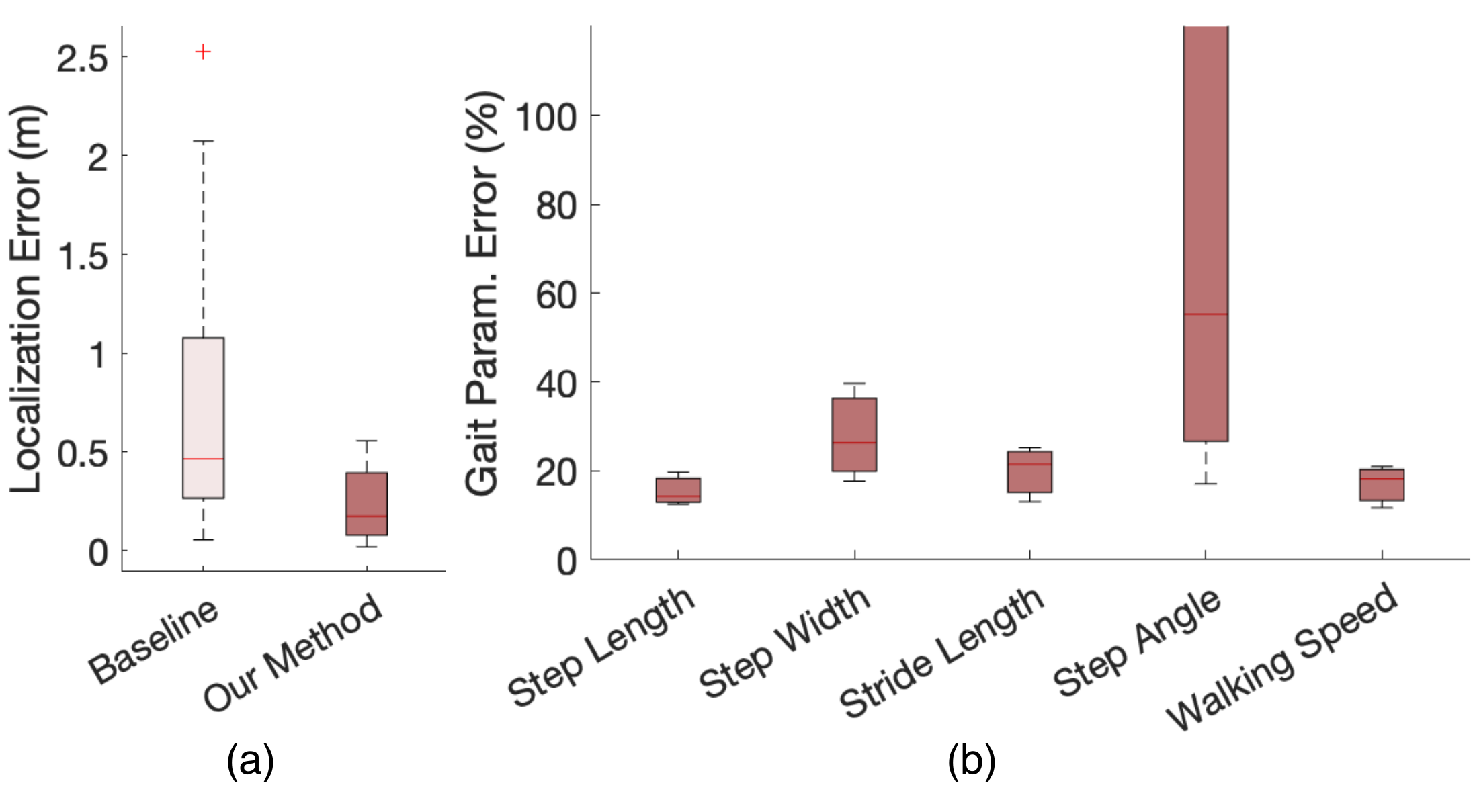}
  \caption{Performance of \name{}: (a) the footstep localization error of our method is 3$\times$ lower than the baseline approach~\cite{mirshekari2018occupant}; (b) Our method estimates the spatial gait parameter with reasonable percentage deviations (27.1\% on average) from their corresponding ground truth values.}%
\label{fig:performance}
\vspace{-10pt}
\end{center}
\end{figure}

\subsubsection{Localization Accuracy}
\name{} has a mean absolute error of 0.22 m for localization, which achieves a 3$\times$ of error reduction as compared to the existing localization approach~\cite{mirshekari2018occupant} with an average error of 0.67 m (see Figure~\ref{fig:performance}a). Note that the error of the baseline is higher than that reported by the original paper, mainly because of the 50$\times$ lower sampling frequency in our data, which is designed for a lower data storage and processing requirement during in-home deployments. In addition, our approach reduces the standard deviation by 2.3$\times$, meaning that the precision of localization is also improved.

\subsubsection{Gait Parameter Estimation Accuracy}
For spatial parameter estimation, \name{} has mean absolute percentage errors (MAPE) of 14\% ($\pm$0.1 m) for step length, 26\% ($\pm$0.05 m) for step width, 21\% ($\pm$0.3 m) for stride length, 55\% ($\pm$1.9$^{\circ}$) for step angle, and 18\% ($\pm$0.21 m/s) for walking speed estimations, respectively. 

Among these results, the error of step angle is significantly larger than the rest of the parameters. This is mainly because the step angles are typically very small when the subject is walking in a straight line without turning in a different direction ($\sim$3.5$^{\circ}$ in our data), leading to a relatively large percentage error. Similar trends are observed when only vibration sensors are used, where a more than 200\% error is observed for step angle estimation. In addition, the step width percentage error is larger than that of the lengths because the average step width is much smaller ($\sim$0.2 m in our data) than the average step length of a given subject ($\sim$0.7 m in our data). 

%% file: Sections/05Conclusion.tex
\section{Future Work}\label{sec:futurework}
For future work, we plan to integrate structural vibration-based sensing with other modalities for gait health monitoring and characterize its sensitivity to the changing environment.

\textbf{Multimodal Fusion for Kinetic and Kinematic Gait Analysis:} While structural vibration contains mainly kinetic gait information, other modalities such as cameras, wearables, and RF sensors capture kinematic gait information~\cite{haque2020illuminating,doheny2010single,chen2017floc}. The integration of both is highly complementary for a more comprehensive, interpretable, and accurate gait analysis. Therefore, we plan to explore further multi-modal fusion approaches that are aware of the reliability of each modality and are able to leverage the advantages of one modality to enhance the other. 

\textbf{Sensitivity Study on Footstep-induced Structural Vibrations:} Structural vibration induced by human footsteps is sensitive to various factors such as structure types/layouts, shoe types, and mental status~\cite{Pan2015}, which may affect the accuracy of gait health monitoring. To characterize and model such influence, we plan to explore the effect of these factors to improve the robustness of our method.

\section{Conclusions}\label{sec:conclusion}
In this study, we develop \name{}, an enhanced vibration-based footstep localization method through multi-modal fusion with temporarily installed cameras for in-home gait analysis. Our system leverages the accurate spatial-temporal information from the temporarily installed cameras to improve the accuracy in vibration-based footstep localization. The main challenge in vibration-based localization is the uncertainty in wave arrival time and the variations in the wave propagation velocity. To address this issue, we characterize such uncertainties and fuse vision and vibration data to model the wave velocity profile and reduce the uncertainty in wave arrival time estimation. We evaluate \name{} through a real-world experiment with 50 walking trials. With 3 trials of multi-modal fusion, we reduce the average localization from 0.67 to 0.22 meters and the spatial parameter estimation error by 4.1$\times$ (from 111.4\% to 27.1\%).